%% file: main.tex
\documentclass[nonacm=true,sigconf]{acmart}
\renewcommand\footnotetextcopyrightpermission[1]{} % removes footnote with conference information in first column
\usepackage{tikz}
\usepackage{mathtools, cuted}
\usepackage{multicol}
\usepackage{svg}
\usepackage{comment}
\usepackage{xcolor}
\usepackage{multirow}
\usepackage{enumitem}
\usepackage{fontawesome}
\usepackage{graphicx}
\usepackage{supertabular}
\usepackage{caption}
\usepackage{subcaption}
\usepackage{amsmath}
\usepackage{amsthm}
\usepackage{bm}
\usepackage{lineno}
\usepackage{tikz}
\usetikzlibrary{positioning, fit, calc}   
\tikzset{block/.style={draw, thick, text width=2cm ,minimum height=1.3cm, align=center},   
line/.style={-latex}     
}  

\setcopyright{rightsretained}
\usepackage{breqn}

\begin{document}

\title{Experimenting with Selected Automated \\ Approaches for Bias Analysis}

\author{Gizem Gezici}
\affiliation{
  \institution{Sabanci University}
  \city{Istanbul}
  \country{Turkey}}
\email{gizem.gezici@sabanciunv.edu}

\renewcommand{\shortauthors}{Gezici, Gizem}

\begin{abstract}

This work first presents our attempts to establish an automated model using state-of-the-art approaches for analysing bias in search results of Bing and Google. Experimental results indicate that the current class-wise F1-scores of our best model are not sufficient to establish an automated model for bias analysis. Thus, we decided not to continue with this approach

%This work suggests an unsupervised pipeline to extract key phrases from textual content. The process takes a given input and detects noun phrases via a rule-based algorithm using information gained from a dependency parser. For selecting the candidate phrases, we leverage sBERT~\cite{sBERT} to represent the noun phrases as well as the original text in the same embedding space. Further, we rank the candidate phrases based on their cosine similarity scores with the original text. The results show that our easily customizable proposed pipeline outperforms the state-of-the-art approaches on the benchmark dataset of NUS \cite{NUS} and gives comparable results on the other three benchmark datasets of INSPEC \cite{INSPEC}, DUC2001 \cite{DUC2001} and SemEval2017\cite{SemEval2017}.
\end{abstract}



\maketitle
\input{body.tex}

\bibliographystyle{ACM-Reference-Format}
\bibliography{main}

\end{document}

%% file: body.tex
\section{Problem Statement}
In our previous work~\cite{gezici2021evaluation}, we achieved to show that there is ideological bias in search results of Bing and Google. It seems that both search engines are biased and they are biased towards the liberal side. Yet, we had fulfilled this study in the top-10 search engine result pages (SERPs) annotated via crowd-sourcing so these results does not help us to investigate the source of bias, i.e. if the bias comes from the dataset, or the ranking algorithm itself. Therefore, as the next step of this research study, we aimed to investigate the source of bias. For detecting the source of bias, we need to annotate the full SERPs of Bing and Google, note that we have 250 web documents for each controversial query and in total there are 57 queries which means that we need an automated model for the annotation phase. 

For these reasons, we firstly extended our annotated dataset and then experimented with different models especially deep learning models to be able to annotate a given document with an acceptable level of accuracy. In the previous report, our models did not give satisfactory results which means that we could not use them for the annotation task, thereby to examine the source of bias in SERPs. Because of this, in this phase of our research we tried various approaches from literature to improve our BERT-based models as well as experimented with more traditional machine learning models such as XGBoost~\cite{Chen2016}, Random Forest~\cite{Liaw2002}.

After we tried different approaches to improve our model, we observed a significant increase in class-wise model evaluation results. However, we believe that class-wise accuracies are still not sufficient for an automated model that will be used for bias analysis. Thus, we left the source of bias analysis part as unsolved.

\section{Automated Bias Analysis in Search Results}
In this section, we aimed to establish an automated model for analysing search results using the state-of-the-art approaches.

\subsection{Correcting the Annotation Dataset \& Obtaining More Labels}
We allocated some time for correcting the labels which were given by crowd-workers. Yet, in this stage we did not touch the annotations for the top-10 SERPs since we already published a journal manuscript that show our findings for the top-10 SERPs. Nonetheless, we experimented with the different versions of the dataset, i.e. removing the incorrect labels, keeping the incorrect labels with the corrected ones. Moreover, we annotated more documents to prevent overfitting and in total we have 3573 labelled documents, i.e. 829 pro instances, 693 agst instances, 1516 neut instances and 535 not-rel instances. For the model training phase, we generated more not-rel instances since our two-phase stance detection model works better when relevant instances exist almost equally to not-relevant instances in the dataset. After experimenting with these different approaches, unfortunately we could not see a sufficient improvement in evaluation results.

\subsection{Examining the Model Errors \& Cleaning the Dataset for Ambigious Instances}
In addition to correcting the labels and annotating more instances, we also examined the model errors. For the instances that the model labels wrongly, we looked at the textual content of the corresponding document and then the probability distribution that the model gave for the given instance. Our main aim here is to detect ambiguous instances which contain different opinions throughout the document. Based on this analysis, we discarded some ambiguous/difficult instances from the dataset and trained our model with this new dataset. Yet, we did not see a big improvement in model results.

\subsection{Experimenting with Traditional Machine Learning Models}
Apart from the BERT-based models, we also experimented with traditional machine learning models, namely Support Vector Machine (SVM)~\cite{cortes1995support}, Random Forest, and XGBoost. Among these, XGBoost gave the best class-wise results as displayed in Table~\ref{tab:ml} but still the model is not sufficiently good to be used for bias analysis.

\begin{table*}[!t]
    \centering
    %\vspace{1em}
    % Engine 1 is better (Bing)
    \caption{F1-scores for SVM, Random Forest, and XGBoost}
    \begin{tabular}{ccccc}
        \hline\hline
        & Pro & Against & Neutral & Not-rel \\
        \hline
        SVM & 0.50 & 0.19 & 0.55 & 0.59 \\
        Random Forest & 0.53 & 0.10 & 0.46 & 0.43 \\
        \textbf{XGBoost} & 0.49 & 0.34 & 0.61 & 0.84 \\
        \hline\hline
    \end{tabular}
    \label{tab:ml}
    %\vspace{1em}
\end{table*}

\subsection{Model Stacking} We also tried the approach of model stacking with the aim of creating a stronger classifier using the weak classifiers that are different from each other as depicted in Fig.~\ref{fig:model_stacking}. This approach slightly improved our results that are not sufficient for bias analysis.

\subsection{Language Model Fine-tuning On Our Stance Dataset} We also applied the steps of Universal Language Model Fine-Tuning (ULMFIT) with the aim of achieving a better domain adaptation. For this, we initially fine-tuned the pre-trained model of BERT on our dataset of query-document content pairs without any stance labels. For this intermediate step, we also experimented with \emph{slanted triangular rates}, \emph{discriminative fine-tuning}, and \emph{gradual unfreezing} which are the proposed approaches in the original ULMFIT paper. Then, using the fine-tuned language model, we fine-tuned it on the stance classification tasks with stance labels. This approach helps to improve the classification performance in the last step on small datasets. However, we could not see a big improvement for our model.

\subsection{Longformer: The Long-Document Transformer} Since transformer-based models, e.g. BERT, RoBERTa, are unable to process long sequences due to their self-attention mechanism, researchers proposed a new transformer to process long sequences. We also experimented with this transformer model and thought that this approach can improve our model results because we have very long documents in our stance dataset. Yet, the results did not show a big improvement.

\subsection{Hyperparameter Tuning \& Applying Mixout}
Apart from the different methods we tried to improve our model results, finally we experimented with different hyperparameters. Despite the strong empirical performance of fine-tuned transformer models, fine-tuning is known to be an unstable process, different random seeds can result in large variance of the task performance. Thus, researchers reported the best hyper-parameter values for BERT, RoBERTa and ALBERT to alleviate the fine-tuning instability. Since we observed fine-tuning instability in our experiments, we experimented with the recommended hyper-parameter values. We observed that these values helped us to achieve a better fine-tuning stability regardless of the random seed, i.e. we obtained similar results with different training runs. However, the results were not sufficient.

\begin{figure*}[!t]
    {\includegraphics[width=0.9\textwidth]{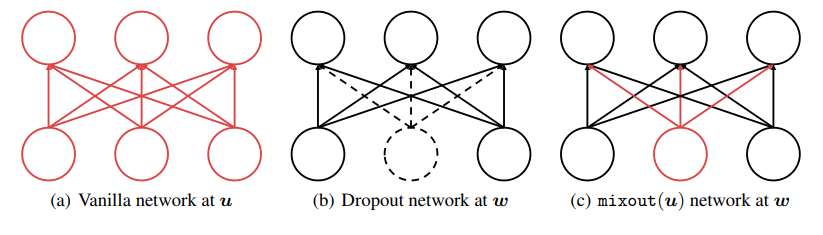}}
    \caption[Model]{The regularisation technique of mixout motivated by dropout}
    \label{fig:Mixout}
\end{figure*}

Lastly, we applied the proposed technique of mixout to regularise the fine-tuning of a pretrained model motivated by another widely-used regularisation technique of dropout, especially for small datasets as displayed in Fig.~\ref{fig:Mixout}. It has been observed that although BERT-large outperforms BERT-base generally, fine-tuning may fail if the target training dataset has less than 10,000 instances. We realised that as mixout helped us to achieve a better fine-tuning stability for transformed-based models and we used mixout with the recommended hyper-parameters. Although, we observed a significant improvement in fine-tuning stability, our models did not give sufficiently good results that can be used for obtaining stance labels. The best results we got using the approaches mentioned above can be seen in Table~\ref{tab:best}. As you can see that in comparison to our previous model results, we achieved better results -- especially the decrease in loss value is quite high but our class-wise accuracies are still not good enough for annotation in bias analysis.

\begin{table*}[!h]
    \centering
    %\vspace{1em}
    % Engine 1 is better (Bing)
    \caption{Best Evaluation Results on the Stance Dataset with BERT-based Model}
    \begin{tabular}{c|cccc|c}
        \hline\hline
        \textbf{Model} & \textbf{Pro} & \textbf{Against} & \textbf{Neutral} & \textbf{Not-rel} & \textbf{Loss}\\
        \hline
        Current Model (More Instances) & 0.58 & 0.76 & 0.52 & 0.93 & 1.52\\
        \hline
        Previous Model & 0.65 & 0.38 & 0.42 & 0.89 & 1.91\\
        \hline\hline
    \end{tabular}
    \label{tab:best}
    %\vspace{1em}
\end{table*}

\begin{figure*}[!t]
    {\includegraphics[width=0.9\textwidth]{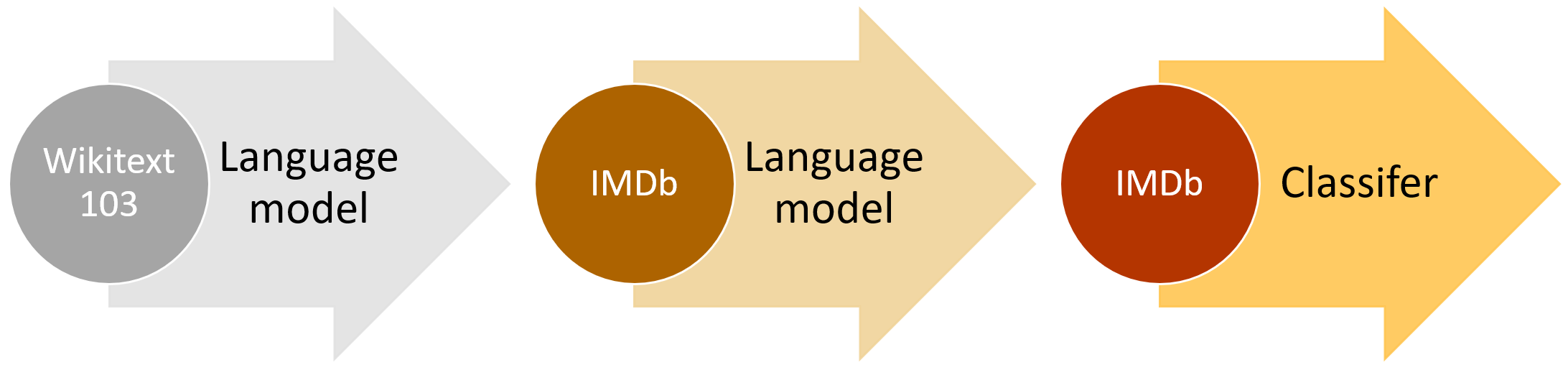}}
    \caption[Model]{Universal Language Model Fine-Tuning (ULMFIT) Steps \footnotemark}
    \label{fig:ULMFIT}
\end{figure*}

\footnotetext{\scriptsize \url{https://yashuseth.blog/2018/06/17/understanding-universal-language-model-fine-tuning-ulmfit/}}